\documentclass[prd,reprint,superscriptaddress,altaffillsymbol,amsmath,nofootinbib]{revtex4-1}
\pdfoutput=1

\usepackage{ifthen}
\usepackage{epsfig}
\usepackage{booktabs} 
\usepackage{natbib}
\usepackage{wasysym}
\usepackage{feynmp}
\usepackage{slashed} 
\usepackage{bbm}
\usepackage{mathrsfs}
\usepackage{amssymb}
\usepackage{dsfont}
\usepackage[export]{adjustbox}

\usepackage{color}

\newcommand{\beqra}{\begin{eqnarray}}
\newcommand{\eeqra}{\end{eqnarray}}
\newcommand{\beq}{\begin{equation}}
\newcommand{\eeq}{\end{equation}}

\begin{document}

\title{Dark matter spin determination with directional direct detection experiments}

\author{Riccardo Catena}
\email{catena@chalmers.se}
\affiliation{Chalmers University of Technology, Department of Physics, SE-412 96 G\"oteborg, Sweden}

\author{Jan Conrad}
\email{conrad@fysik.su.se}
\affiliation{Oskar Klein Centre, Department of Physics, Stockholm University, AlbaNova, Stockholm SE-10691, Sweden}

\author{Christian D\"oring}
\email{cdoering@mpi-hd.mpg.de}
\affiliation{Institut f\"ur Theoretische Physik, Friedrich-Hund-Platz 1, 37077 G\"ottingen, Germany}
\affiliation{Max-Planck-Institut f\"ur Kernphysik, 69117 Heidelberg, Germany}

\author{Alfredo Davide Ferella}
\email{alfredo.ferella@fysik.su.se}
\affiliation{Oskar Klein Centre, Department of Physics, Stockholm University, AlbaNova, Stockholm SE-10691, Sweden}

\author{Martin B.~Krauss}
\email{martin.krauss@chalmers.se}
\affiliation{Chalmers University of Technology, Department of Physics, SE-412 96 G\"oteborg, Sweden}

\begin{abstract}If the dark matter particle has spin 0, only two types of WIMP-nucleon interaction can arise from the non-relativistic reduction of renormalisable single-mediator models for dark matter-quark interactions.~Based on this crucial observation, we show that about 100 signal events at next generation directional detection experiments can be enough to enable a $2\sigma$ rejection of the spin 0 dark matter hypothesis in favour of alternative hypotheses where the dark matter particle has spin 1/2 or 1.~In this context directional sensitivity is crucial, since anisotropy patterns in the sphere of nuclear recoil directions depend on the spin of the dark matter particle.~For comparison, about 100 signal events are expected in a CF$_4$ detector operating at a pressure of 30 torr with an exposure of approximately 26,000 cubic-meter-detector days for WIMPs of 100 GeV mass and a WIMP-Fluorine scattering cross-section of 0.25~pb.~Comparable exposures are within reach of an array of cubic meter time projection chamber detectors.
\end{abstract}


\maketitle

\section{Introduction}
\label{sec:intro}
Increasingly accurate cosmological observations show that our Universe's energy budget is presently dominated by dark energy and dark matter, representing 69\% and 26\% of the total energy content, respectively~\cite{Ade:2015xua}.~In the standard paradigm of modern Cosmology, the dark matter component of the Universe is made of new hypothetical particles which have so far escaped detection~\cite{Jungman:1995df}.~The detection of dark matter particles from the Cosmos is arguably the most pressing question in astroparticle physics today.~If dark matter is made of Weakly Interacting Massive Particles (WIMPs), the experimental technique known as direct detection will be pivotal in this context.~It searches for nuclear recoil events induced by the scattering of Milky Way dark matter particles in low-background detectors.~The field of dark matter direct detection has progressed very rapidly in recent years.~The XENON1T experiment has recently released its first data, setting the most stringent exclusion limits on the spin-independent WIMP-nucleon scattering cross-section, $\sigma_{\rm SI}$, for WIMP masses above 10 GeV, with a minimum of $7.7\times10^{-47}$~cm$^{2}$ for 35 GeV WIMPs at 90\% confidence level~\cite{Aprile:2017iyp}.~This result improves previous limits on $\sigma_{\rm SI}$ in the same WIMP mass range reported by the LUX~\cite{Akerib:2016vxi} and PANDAX-II~\cite{Tan:2016zwf} experiments.~Substantial progress has also been made in the search for light WIMPs.~CRESST-II~\cite{Angloher:2015ewa} has been operating with an energy threshold for nuclear recoils of 307 eV, deriving the most stringent limits on $\sigma_{\rm SI}$ in the WIMP mass region below 1.8 GeV.~In the mass range around 5 GeV, the most stringent limits on $\sigma_{\rm SI}$ have been set by CDMS-lite~\cite{Agnese:2015nto}.~Finally, the best direct-detection constraints on spin-dependent WIMP-proton and WIMP-neutron scattering cross-sections have been derived by PICO-60~\cite{Amole:2017dex} and LUX~\cite{Akerib:2017kat}, respectively.

In the next decade, complementary strategies will be pursued in order to improve current detection methods, and thus achieve the first WIMP direct detection.~A first strategy consists in increasing the target mass of current detectors.~Experiments that will focus on this strategy include XENONnT~\cite{Aprile:2015uzo}, LZ~\cite{Mount:2017qzi} and DARWIN~\cite{Aalbers:2016jon}.~These experiments will operate in the coming years exploiting multi-ton double phase Xenon detectors.~A second strategy consists in lowering the experimental energy threshold in order to gain sensitivity to dark matter particles in the sub-GeV mass range.~This approach will be explored by, e.g., CRESST-III~\cite{Strauss:2016sxp} and SuperCDMS~\cite{Agnese:2016cpb}.~A third possibility is to focus on the time dependence of the rate of nuclear recoil events at detector.~For example, an annual modulation in the rate of dark matter-induced nuclear recoil events is expected because of the inclination of the Earth's orbit with respect to the galactic plane.~Furthermore, the Earth-crossing of WIMPs can induce a distinctive daily modulation in the number of signal events~\cite{Kavanagh:2016pyr}.~Anais~\cite{Amare:2015qcn}, Cosinus~\cite{Angloher:2016ooq}, DM-Ice~\cite{deSouza:2016fxg} and Sabre~\cite{Froborg:2016ova}, will explore this possibility with the goal of validating or conclusively rule out the dark matter interpretation of the DAMA modulation signal.~This signal is currently at odds with a variety of direct detection experiments, despite its statistical significance being at the 9$\sigma$ level, if standard assumptions regarding astro-, particle and nuclear physics are made~\cite{Bernabei:2010mq} (see however~\cite{Catena:2016hoj} for a critical reassessment of DAMA's results).~Finally, a fourth strategy consists in developing detectors that are sensitive to the direction of nuclear recoils induced by the scattering of dark matter particles in gaseous target materials~\cite{Mayet:2016zxu}.~Efforts in developing detectors with directional sensitivity are well-motivated, since the expected angular distribution of dark matter-induced nuclear recoil events is not isotropic, as the Earth's motion selects a preferred direction in the sphere of recoil directions~\cite{Spergel:1987kx}.~Among directional detection experiments currently in a research and development stage are DRIFT~\cite{Battat:2014van,Daw:2011wq}, MIMAC~\cite{Riffard:2013psa,Santos:2013hpa}, DMTPC~\cite{Battat:2012,Monroe:2012qma}, NEWAGE~\cite{Miuchi:2012rma,Miuchi:2010hn} and D3~\cite{Vahsen:2011qx}.~Typically, diffuse gas detectors and time projection chambers are used to reconstruct the nuclear recoil tracks, but alternative approaches are also considered.~These include the use of nuclear emulsions~\cite{Naka:2011sf}, dark matter-electron scattering in crystals~\cite{Essig:2011nj} and DNA detectors~\cite{Drukier:2012hj}.

The purpose of this work is to show that directional detectors can be used to constrain the dark matter particle spin.~Directional detectors can be used for ``WIMP spin model selection'' as a result of recent theoretical developments that we now briefly review.~Recoil energy spectra and angular distributions expected at dark matter directional detectors have in the past been calculated for two types of dark matter-nucleon interaction only:~the so-called spin-independent and spin-dependent interactions.~Recently, this calculation has been extended to the full set of non-relativistic operators for dark matter-nucleon interactions that are compatible with Galilean invariance, and that are at most linear in the transverse relative velocity operator~\cite{Catena:2015vpa,Kavanagh:2015jma}  -- a framework also known as non-relativistic effective theory of dark matter-nucleon interactions~\cite{Fan:2010gt,Fitzpatrick:2012ix,DelNobile:2013sia,Catena:2014uqa,Catena:2014hla,Catena:2014epa,Catena:2015uha,Catena:2015uua,Catena:2016kro}.~For later convenience, the sixteen Galilean invariant operators predicted by the effective theory of dark matter-nucleon interactions are listed in Tab.~\ref{tab:operators}.~The extension of the standard paradigm based upon the canonical spin-independent and spin-dependent interactions to the non-relativistic effective theory of dark matter-nucleon interaction has led to the discovery of new potentially important signatures of particle dark matter.~The most striking result found in~\cite{Catena:2015vpa,Kavanagh:2015jma} is that the angular distribution of nuclear recoil events generated by the interaction operators $\hat{\mathcal{O}}_{5}$, $\hat{\mathcal{O}}_{7}$, $\hat{\mathcal{O}}_{8}$, $\hat{\mathcal{O}}_{13}$ and $\hat{\mathcal{O}}_{14}$ in Tab.~\ref{tab:operators} has a maximum in rings centred around the direction of the Earth's motion in the galactic rest frame, $\mathbf{v}_{\oplus}$.~For the remaining interaction operators in Tab.~\ref{tab:operators}, recoil events are mainly expected in the direction opposite to $\mathbf{v}_{\oplus}$.~A second theoretical development that we will exploit in the present analysis is the recent systematic classification and characterisation of the non-relativistic limit of so-called ``simplified models'' -- single-mediator models for dark matter.~In~\cite{Dent:2015zpa} it has been shown that the Galilean invariant operators in Tab.~\ref{tab:operators} arise from the non-relativistic reduction of renormalisable single-mediator models for dark matter-quark interactions, although not all of them as leading operators~\cite{Bishara:2016hek,Hoferichter:2015ipa}.~In particular, within the framework of \cite{Dent:2015zpa} it is possible to predict the subset of operators in Tab.~\ref{tab:operators} that can be associated to a given dark matter particle spin.~As a result, the link between models in~\cite{Dent:2015zpa} and non-relativistic operators in Tab.~\ref{tab:operators} establishes a correspondence between dark matter spin and ring-like features in the sphere of recoil directions.~This correspondence is the theoretical input that we propose to use to constrain the WIMP spin.

\begin{table}[t]
    \centering
    \begin{ruledtabular}	
    \begin{tabular}{l}
    \toprule
        $\hat{\mathcal{O}}_1 = \mathds{1}_{\chi}\mathds{1}_N$  \\  
        $\hat{\mathcal{O}}_3 = i{\bf{\hat{S}}}_N\cdot\left(\frac{{\bf{\hat{q}}}}{m_N}\times{\bf{\hat{v}}}^{\perp}\right)\mathds{1}_\chi$ \\
        $\hat{\mathcal{O}}_4 = {\bf{\hat{S}}}_{\chi}\cdot {\bf{\hat{S}}}_{N}$ \\                                                      
        $\hat{\mathcal{O}}_5 = i{\bf{\hat{S}}}_\chi\cdot\left(\frac{{\bf{\hat{q}}}}{m_N}\times{\bf{\hat{v}}}^{\perp}\right)\mathds{1}_N$ \\       
        $\hat{\mathcal{O}}_6 = \left({\bf{\hat{S}}}_\chi\cdot\frac{{\bf{\hat{q}}}}{m_N}\right) \left({\bf{\hat{S}}}_N\cdot\frac{\hat{{\bf{q}}}}{m_N}\right)$ \\  
        $\hat{\mathcal{O}}_7 = {\bf{\hat{S}}}_{N}\cdot {\bf{\hat{v}}}^{\perp}\mathds{1}_\chi$ \\ 
        $\hat{\mathcal{O}}_8 = {\bf{\hat{S}}}_{\chi}\cdot {\bf{\hat{v}}}^{\perp}\mathds{1}_N$  \\ 
        $\hat{\mathcal{O}}_9 = i{\bf{\hat{S}}}_\chi\cdot\left({\bf{\hat{S}}}_N\times\frac{{\bf{\hat{q}}}}{m_N}\right)$ \\ 
        $\hat{\mathcal{O}}_{10} = i{\bf{\hat{S}}}_N\cdot\frac{{\bf{\hat{q}}}}{m_N}\mathds{1}_\chi$   \\
        $\hat{\mathcal{O}}_{11} = i{\bf{\hat{S}}}_\chi\cdot\frac{{\bf{\hat{q}}}}{m_N}\mathds{1}_N$   \\
        $\hat{\mathcal{O}}_{12} = {\bf{\hat{S}}}_{\chi}\cdot \left({\bf{\hat{S}}}_{N} \times{\bf{\hat{v}}}^{\perp} \right)$  \\    
        $\hat{\mathcal{O}}_{13} =i \left({\bf{\hat{S}}}_{\chi}\cdot {\bf{\hat{v}}}^{\perp}\right)\left({\bf{\hat{S}}}_{N}\cdot \frac{{\bf{\hat{q}}}}{m_N}\right)$  \\ 
        $\hat{\mathcal{O}}_{14} = i\left({\bf{\hat{S}}}_{\chi}\cdot \frac{{\bf{\hat{q}}}}{m_N}\right)\left({\bf{\hat{S}}}_{N}\cdot {\bf{\hat{v}}}^{\perp}\right)$  \\          
        $\hat{\mathcal{O}}_{15} = -\left({\bf{\hat{S}}}_{\chi}\cdot \frac{{\bf{\hat{q}}}}{m_N}\right)\left[ \left({\bf{\hat{S}}}_{N}\times {\bf{\hat{v}}}^{\perp} \right) \cdot \frac{{\bf{\hat{q}}}}{m_N}\right] $    \\ 
        $\hat{\mathcal{O}}_{17}=i \frac{{\bf{\hat{q}}}}{m_N} \cdot \mathbf{\mathcal{S}} \cdot {\bf{\hat{v}}}^{\perp} \mathds{1}_N$ \\          
$\hat{\mathcal{O}}_{18}=i \frac{{\bf{\hat{q}}}}{m_N} \cdot \mathbf{\mathcal{S}}  \cdot {\bf{\hat{S}}}_{N}$ \\                                                                     
    \bottomrule
    \end{tabular}
    \end{ruledtabular}	
    \caption{Quantum mechanical operators defining the non-relativistic effective theory of dark matter-nucleon interactions~\cite{Fan:2010gt,Fitzpatrick:2012ix}.~The notation is the one introduced in Sec.~\ref{sec:theory}.~The operators $\hat{\mathcal{O}}_{17}$ and $\hat{\mathcal{O}}_{18}$ only arise for spin 1 WIMPs, and $\mathbf{\mathcal{S}}$ is a symmetric combination of spin 1 WIMP polarisation vectors~\cite{Dent:2015zpa}.} 
    \label{tab:operators}
\end{table}

The method for WIMP spin model selection that we will develop in this work can in principle be applied to arbitrary spin configurations.~However, in this work we will focus on the prospects for rejecting the spin 0 WIMP hypothesis.~In this case, our approach to WIMP spin model selection can be illustrated as follows:~the interaction operators $\hat{\mathcal{O}}_{1}$ and $\hat{\mathcal{O}}_{10}$ are the only ones that can arise from the non-relativistic reduction of renormalisable single-mediator models for spin 0 WIMPs~\cite{Dent:2015zpa}.~Consequently, rejecting the spin 0 WIMP hypothesis in favour of other WIMP spin configurations is equivalent to rejecting the operators $\hat{\mathcal{O}}_{1}$ and $\hat{\mathcal{O}}_{10}$ in favour of other interactions.~Here we argue that directional detection experiments can be used for this purpose.~Indeed, $\hat{\mathcal{O}}_{1}$ and $\hat{\mathcal{O}}_{10}$ are characterised by angular distributions of nuclear recoil events which do not exhibit ring-like patterns.~Therefore, the rejection of $\hat{\mathcal{O}}_{1}$ and $\hat{\mathcal{O}}_{10}$ can be based upon the search for ring-like features in the sphere of dark matter-induced nuclear recoil events.~In this work, we will show that about 100 signal events at next generation directional detection experiments can be enough to enable a $2\sigma$ rejection of the spin 0 WIMP hypothesis in favour of alternative hypotheses where the dark matter particle has spin 1/2 or 1.

The paper is organised as follows.~In Sec.~\ref{sec:theory}, we review the framework used to model WIMP-quark and -nucleon interactions.~In Sec.~\ref{sec:met}, we introduce our approach to WIMP spin model selection, and explain how experimental data on nuclear recoil energies and directions have been simulated in order to validate our proposal.~Sec.~\ref{sec:results} is devoted to our results, while in Sec.~\ref{sec:conclusion} we summarise and conclude.

\section{Theoretical framework}
\label{sec:theory}
In this section we review the theoretical framework introduced in~\cite{Dent:2015zpa} and used here to model the interactions of dark matter with quarks and nucleons.~In reviewing~\cite{Dent:2015zpa}, we show that if dark matter has spin 0, only two types of WIMP-nucleon interaction can be generated in the non-relativistic limit of renormalisable single-mediator models for dark matter-quark interactions.~The corresponding interaction operators are denoted by $\hat{\mathcal{O}}_1$ and $\hat{\mathcal{O}}_{10}$ in Tab.~\ref{tab:operators}.~Because of this crucial property of spin 0 dark matter, rejecting the hypothesis of a spin 0 WIMP is equivalent to rejecting $\hat{\mathcal{O}}_1$ and $\hat{\mathcal{O}}_{10}$ in favour of alternative interaction operators.~In contrast, if dark matter has spin 1/2 or 1, a variety of WIMP-nucleon interactions can arise, and rejecting these spin configurations is generically more difficult.~Interestingly, there are WIMP-nucleon interactions that are specific to spin 1 or spin 1/2 WIMPs, as we will see in Secs.~\ref{sec:spin1} and~\ref{sec:spin1/2}, respectively. 

\subsection{Spin 1 dark matter}
\label{sec:spin1}
Spin 1 WIMPs can interact with quarks through the exchange of spin 0, 1/2 or 1 particles.~Here we focus on the Lagrangian
\begin{align}
\mathcal{L}^{(1)}_{\rm int} = 
-b_6\,\partial_\mu \left( X^{\mu\dagger} X_\nu + X_\nu^\dagger X^\mu \right) G^\nu - h_3\sum_q \bar{q} \gamma^\mu q G_\mu  \,,
\label{eq:Lspin1}
\end{align}
which describes possible interactions of a spin 1 WIMP, $X_\mu$, with the Standard Model quarks, $q$.~Interactions are mediated by the vector boson $G_\nu$.~Model parameters are the real coupling constants $b_6$ and $h_3$, and the dark matter and mediator mass, $m_X$ and $m_G$, respectively.~The cross-section for WIMP-nucleus scattering can be computed from the WIMP-nucleon scattering amplitude.~In the non-relativistic limit, Eq.~(\ref{eq:Lspin1}) predicts the following amplitude for WIMP-nucleon scattering
\begin{align}
\mathcal{M}^{(1)}_{\rm NR} &= - \frac{3 h_3 b_6}{m_G^2} \frac{m_N}{m_X} \left[ \hat{\mathbf{S}}_X^{r'r} \cdot \left( \frac{i \mathbf{q}}{m_N} \times \mathbf{v}^\perp \right) \xi^{s' \dagger} \mathds{1}_N \xi^s  \right.  \nonumber\\ 
&\left.+ \left(\frac{\mathbf{q}}{m_N} \cdot \hat{\mathbf{S}}_X^{r'r} \right) \left( \frac{\mathbf{q}}{m_N} \cdot \xi^{s' \dagger} \hat{\mathbf{S}}_N  \xi^s \right)  \right.  \nonumber\\ 
&\left. -\frac{q^2}{m_N^2}  \hat{\mathbf{S}}_X^{r'r} \cdot \xi^{s' \dagger} \hat{\mathbf{S}}_N  \xi^s \right] \,,
\label{eq:Mspin1}
\end{align}
where $(\hat{\mathbf{S}}_{X}^{r'r})_k\equiv \langle r' | (\hat{\mathbf{S}}_{X})_k | r \rangle = -i \epsilon_{ijk} \,\varepsilon_i^{r'*} \varepsilon_j^r$, $(\hat{\mathbf{S}}_{X})_k$, $k=1,2,3$, is the $k$-th component of the dark matter particle spin operator, the kets $|r\rangle$ and $|r'\rangle$ represent initial and final polarisation states, respectively, and $\varepsilon_j^r=\delta_j^r$, $r=1,2,3$ are polarisation vectors.~Two-component Pauli spinors are denoted by $\xi^s$, whereas the matrices $\mathds{1}_N$ and $\hat{\mathbf{S}}_N=\boldsymbol{\sigma}/2$ are the $2\times 2$ identity and nucleon spin operator, respectively.~Here the vector $\boldsymbol{\sigma}$ represents the three Pauli matrices and $m_N$ is the nucleon mass.~Finally, $\mathbf{q}=\mathbf{k}-\mathbf{k}'$ is the momentum transferred, $\mathbf{k}$ and $\mathbf{k}'$ are initial and final nucleon three-momenta, and $\mathbf{v}^\perp$ is the WIMP-nucleon transverse relative velocity.~By construction, $\mathbf{v}^\perp \cdot \mathbf{q}=0$~\cite{Fitzpatrick:2012ix}.~Eq.~(\ref{eq:Lspin1}) implies isoscalar WIMP-nucleon interactions, i.e.~the same coupling to protons and neutrons.~In the first order Born approximation, $\mathcal{M}^{(1)}_{\rm NR}$ is also given by 
\begin{align}
\mathcal{M}^{(1)}_{\rm NR} = - \langle N(\mathbf{k}',s')| \,\mathcal{V}_{r'r}^{(1)}\,|N(\mathbf{k},s) \rangle \,,
\label{eq:Bspin1}
\end{align}
where $|N(\mathbf{k},s )\rangle$ represents a single-nucleon state and $\mathcal{V}^{(1)}_{r'r}$ is the non-relativistic quantum mechanical potential
\begin{align}
\mathcal{V}^{(1)}_{r'r}&= \frac{3 h_3 b_6}{m_G^2} \frac{m_N}{m_X} \left[ \hat{\mathbf{S}}_X^{r'r} \cdot \left(\frac{i \hat{\mathbf{q}}}{m_N} \times \hat{\mathbf{v}}^\perp \right) \mathds{1}_N  \right.  \nonumber\\
&\left.+ \left( \frac{i \hat{\mathbf{q}}}{m_N} \cdot \hat{\mathbf{S}}_X^{r'r} \right) \left( \frac{i \hat{\mathbf{q}}}{m_N} \cdot \hat{\mathbf{S}}_N \right)  \right.  \nonumber\\
&\left. -\frac{q^2}{m_N^2}  \hat{\mathbf{S}}_X^{r'r} \cdot \hat{\mathbf{S}}_N  \right] \,.
\label{eq:Vspin1}
\end{align}
In Eq.~(\ref{eq:Vspin1}), $\hat{\mathbf{q}}$ and $\hat{\mathbf{v}}^\perp$ are Hermitian and Galilean invariant operators acting on single-nucleon states $|N(\mathbf{k},s )\rangle$.~Introducing the notation:~$\hat{\mathcal{O}}_4=\hat{\mathbf{S}}_X \cdot \hat{\mathbf{S}}_N  $, $\hat{\mathcal{O}}_5=\hat{\mathbf{S}}_X \cdot (i \hat{\mathbf{q}}/m_N \times \hat{\mathbf{v}}^\perp ) \mathds{1}_N$, and $\hat{\mathcal{O}}_6=( \hat{\mathbf{S}}_X \cdot  i \hat{\mathbf{q}}/m_N) ( \hat{\mathbf{S}}_N \cdot i \hat{\mathbf{q}}/m_N) $, the potential $\mathcal{V}^{(1)}_{r r'}$ in Eq.~(\ref{eq:Vspin1}) can be rewritten in the notation of~\cite{Fitzpatrick:2012ib}.~Neglecting two-nucleon currents, the amplitude for WIMP-nucleus scattering is given by 
\begin{align}
\mathcal{M}^{(1)}_{{\rm NR};A} = - \sum_{j=1}^A\int {\rm d}\mathbf{r} \, e^{- i \mathbf{q}\cdot \mathbf{r}} \, \langle \,f\, | \,\mathcal{V}_{r'r}^{(1)}(j)\,|\,i\, \rangle \,,
\label{eq:BNspin1}
\end{align} 
where $A$ is the nucleus mass number, $\mathbf{r}$ the WIMP-nucleus centre of mass relative distance, and $|\,i\, \rangle$ and $|\,f\, \rangle$ initial and final nuclear state, respectively.~In Eq.~(\ref{eq:BNspin1}), the $j$-th nucleon contributes to the amplitude $\mathcal{M}^{(1)}_{{\rm NR};A}$ through the potential $\mathcal{V}_{r'r}^{(1)}(j)$, which is equal to $\mathcal{V}_{r'r}^{(1)}$ but with $\mathbf{S}_{N}$ and $\mathds{1}_N$ now defined in the $j$-th nucleon spin space, and with $\hat{\mathbf{q}}$ and $\mathbf{v}^\perp$ decomposed into a term acting on the nucleus centre of mass coordinates and a term acting on the internal nucleon coordinates~\cite{Fitzpatrick:2012ib}.~Corrections to the WIMP-nucleus scattering amplitude due to two-nucleon currents have only been computed for spin 1/2 dark matter, and for selected nuclear currents~\cite{Klos:2013rwa}.~In this case, it has been found that two-nucleon currents can be important in the low-momentum transfer limit.~Specifically, for odd-neutron (odd-proton) nuclei, two-nucleon currents can significantly increase the WIMP-proton (WIMP-neutron) scattering cross-section, because of strong interactions between nucleons arising from meson exchange~\cite{Hoferichter:2015ipa}.~To extend these calculations to arbitrary WIMP spins and nuclear currents goes beyond the scope of the present work.~Finally, the differential cross-section for WIMP-nucleus scattering is given by
\begin{align}
\frac{{\rm d}\sigma^{(1)}}{{\rm d}E_R} = \frac{2 m_A}{12\pi v^2(2J+1)} \sum_{\rm spins} \big|\mathcal{M}^{(1)}_{{\rm NR};A} \big|^2\,,
\label{eq:sigma1}
\end{align}
where $J$ and $m_A$ are the spin and mass of the target nucleus, respectively, $E_R$ is the nuclear recoil energy, and the sum runs over initial and final spin/polarisation states.~In the numerical applications, $|\mathcal{M}^{(1)}_{{\rm NR};A}|^2$ is expanded in nuclear response functions which are quadratic in matrix elements of nuclear charges and currents, as shown in~\cite{Fitzpatrick:2012ib}.~Notice that the interaction operator $\hat{\mathcal{O}}_5$, i.e.~the first term in Eq.~(\ref{eq:Bspin1}), gives the leading contribution to Eq.~(\ref{eq:sigma1}).~Eq.~(\ref{eq:Lspin1}) is the only renormalisable single-mediator Lagrangian that can generate $\hat{\mathcal{O}}_5$ as a leading operator in the non-relativistic limit.~Therefore, the interaction operator $\hat{\mathcal{O}}_5$ is specific to spin 1 WIMPs.~We will refer to a scenario where dark matter has spin 1 and the potential responsible for WIMP-nucleon interactions is $\mathcal{V}^{(1)}_{r'r}= c_5 \langle r' | \hat{\mathcal{O}}_5 | r \rangle$, $c_5\in\mathbb{R}$, as ``WIMP spin hypothesis $\mathscr{H}_5^{s=1}$".

\subsection{Spin 1/2 dark matter}
\label{sec:spin1/2}
Spin 1/2 WIMPs can interact with quarks through the exchange of spin 0 and spin 1 mediators only.~Spin 1/2 mediators are not allowed by Lorentz invariance.~Here we focus on the interaction Lagrangian
\begin{align}
\mathcal{L}^{(1/2)}_{\rm int} =  - \lambda_3 \,\bar{\chi} \gamma^\mu  \chi G_\mu   - h_4 \sum_q \bar{q} \gamma^\mu \gamma^5 q  G_\mu \,,
\label{eq:Lspin1/2}
\end{align}
which describes the interactions of a WIMP $\chi$ of spin 1/2 and mass $m_\chi$.~The mediator is a spin 1 particle $G_\nu$ of mass $m_G$, and $h_4$,$\lambda_3\in\mathbb{R}$.~From Eq.~(\ref{eq:Lspin1/2}) one obtains the non-relativistic potential 
\begin{align}
\mathcal{V}^{(1/2)}_{r'r}&= \frac{2 \Delta h_4 \lambda_3}{m_G^2} \left( \eta^{r'\dagger} \hat{\mathcal{O}}_7 \eta^{r} - \frac{m_N}{m_\chi} \eta^{r'\dagger} \hat{\mathcal{O}}_9\eta^{r} \right) \,,
\label{eq:Vspin1/2}
\end{align}
where $\Delta=0.33$~\cite{Dent:2015zpa}, $\eta^r$ and $\eta^{r'\dagger}$ are two-component Pauli spinors for the spin 1/2 WIMP $\chi$, $\hat{\mathcal{O}}_7=\hat{\mathbf{S}}_N \cdot \hat{\mathbf{v}}^\perp\mathds{1}_\chi$, $\hat{\mathcal{O}}_9=\hat{\mathbf{S}}_\chi \cdot ( \hat{\mathbf{S}}_N \times i \hat{\mathbf{q}}/m_N)$, $\mathds{1}_\chi$ denotes the identity in the $\chi$ spin space, and $\hat{\mathbf{S}}_\chi=\boldsymbol{\sigma}/2$ is the $\chi$ spin operator.~The potential $\mathcal{V}^{(1/2)}_{r'r}$ allows to calculate the amplitude for WIMP-nucleus scattering, $\mathcal{M}^{(1/2)}_{{\rm NR};A}$, through an equation analogous to Eq.~(\ref{eq:BNspin1}).~The WIMP-nucleus scattering cross-section is now given by
\begin{align}
\frac{{\rm d}\sigma^{(1/2)}}{{\rm d}E_R} = \frac{2 m_A}{8\pi v^2(2J+1)} \sum_{\rm spins} \big|\mathcal{M}^{(1/2)}_{{\rm NR};A} \big|^2\,.
\label{eq:sigma1/2}
\end{align}
Eq.~(\ref{eq:Lspin1/2}) is the only renormalisable single-mediator Lagrangian that can generate $\hat{\mathcal{O}}_7$.~Numerically we find that $\hat{\mathcal{O}}_7$ is the leading operator in Eq.~(\ref{eq:sigma1/2}) in the dark matter particle mass range around 50 GeV.~We will refer to a scenario where dark matter has spin 1/2 and the potential responsible for WIMP-nucleon interactions is $\mathcal{V}^{(1/2)}_{r'r}= c_7 \,\eta^{r'\dagger}\hat{\mathcal{O}}_7 \eta^{r}$, $c_7\in\mathbb{R}$, as ``WIMP spin hypothesis $\mathscr{H}_7^{s=1/2}$". 

\subsection{Spin 0 dark matter}
The most general renormalisable Lagrangian for scalar mediation of spin 0 WIMP interactions with quarks is given by
\begin{align}
\label{eq:phi}
\mathcal{L}^{(0)}_{\rm int;\phi} &= -g_1m_SS^\dag S\phi -\frac{g_2}{2}S^\dag S\phi^2-h_1\bar{q} q\phi-ih_2\bar{q}\gamma^5q\phi \nonumber \\
& -\frac{\lambda_S}{2}(S^{\dag}{S})^2 -\frac{m_\phi\mu_1}{3}\phi^3-\frac{\mu_2}{4}\phi^4  \,,
\end{align}
where $S$ is a complex scalar field describing a stable spin 0 WIMP and $\phi$ is a real scalar mediating the WIMP-quark interaction.~The coupling constants $g_1$, $g_2$, $h_1$, $h_2$, $\lambda_S$, $\mu_1$ and $\mu_2$ are assumed to be real.~We denote by $m_S$ and $m_\phi$ the dark matter particle and mediator mass, respectively.~The most general renormalisable Lagrangian for vector mediation of spin 0 WIMP interactions with quarks is given by
\begin{align}
\label{eq:G}
\mathcal{L}^{(0)}_{\rm int; G_\nu} &= -\frac{g_3}{2}S^{\dag}SG_{\mu}G^{\mu} -ig_4(S^{\dag}\partial_{\mu}S-\partial_{\mu}S^{\dag}S)G^{\mu}\nonumber\\
&-h_3(\bar{q}\gamma_{\mu}q)G^{\mu}-h_4(\bar{q}\gamma_{\mu}\gamma^5q)G^{\mu} \nonumber\\
&-\frac{\lambda_S}{2}(S^{\dag}{S})^2 -\frac{\lambda_G}{4}(G_{\mu}G^{\mu})^2 \,,
\end{align}
where the notation is the same as above, but now the WIMP-quark interaction is mediated by a vector field $G_\mu$ of mass $m_G$.~The coupling constants $g_3$, $g_4$, $h_3$, $h_4$ and $\lambda_G$ are assumed to be real.  

Despite the variety of terms appearing in Eqs.~(\ref{eq:phi}) and (\ref{eq:G}), in the non-relativistic limit only two WIMP-nucleon interaction operators can arise if dark matter has spin 0.~The scattering of spin 0 WIMPs by free nucleons is therefore fully characterised by the following non-relativistic potential
\begin{align}
\mathcal{V}^{(0)} = \left( c_1\,\hat{\mathcal{O}}_1 + c_{10}\,\hat{\mathcal{O}}_{10} \right) \,,
\label{eq:Vspin0}
\end{align}
where $\hat{\mathcal{O}}_1=\mathds{1}_\chi\mathds{1}_N$ and $\hat{\mathcal{O}}_{10}=i\hat{\mathbf{q}}/m_N\cdot\hat{\mathbf{S}}_N\mathds{1}_\chi$.~For spin 0 dark matter, $\mathds{1}_\chi$ is simply equal to 1.~The differential cross-section for WIMP-nucleus scattering is now given by
\begin{align}
\frac{{\rm d}\sigma^{(0)}}{{\rm d}E_R} = \frac{2 m_A}{4\pi v^2(2J+1)} \sum_{\rm spins} \big|\mathcal{M}^{(0)}_{{\rm NR};A}|^2 
\label{eq:sigma0}
\end{align}
where the amplitude $\mathcal{M}^{(0)}_{{\rm NR};A}$ is obtained by inserting the potential $\mathcal{V}^{(0)}$ into an equation analogous to Eq.~(\ref{eq:BNspin1}).~The coupling constants $c_1$ and $c_{10}$ in Eq.~(\ref{eq:Vspin0}) can be expressed in terms of the coupling constants in Eqs.~(\ref{eq:phi}) and (\ref{eq:G}).~We will refer to a scenario where dark matter has spin 0 and the potential responsible for WIMP-nucleon interactions is the one in Eq.~(\ref{eq:Vspin0}) as ``WIMP spin hypothesis $\mathscr{H}_1^{s=0}$'', if $c_{10}=0$, and as ``WIMP spin hypothesis $\mathscr{H}_{10}^{s=0}$'', if $c_1=0$.~For simplicity, in this investigation we will neglect the case where $c_1\neq0$ and $c_{10}\neq0$ simultaneously.~For later convenience, the four WIMP spin hypotheses introduced in this section are summarised in Tab.~\ref{tab:hyp}.
\begin{table}[t]
    \centering
    \begin{ruledtabular}	
    \begin{tabular}{ccc}
    Hypothesis & WIMP spin & Interaction operator \\
    \toprule
   $\mathscr{H}_1^{s=0}$  & $0$ & $\hat{\mathcal{O}}_1$  \\
   $\mathscr{H}_{10}^{s=0}$  & $0$ & $\hat{\mathcal{O}}_{10}$  \\
   $\mathscr{H}_7^{s=1/2}$  & $1/2$ & $\hat{\mathcal{O}}_7$  \\
   $\mathscr{H}_5^{s=1}$  & $1$ & $\hat{\mathcal{O}}_5$  \\	
    \bottomrule   
    \end{tabular}
    \end{ruledtabular}	
    \caption{Summary of the four WIMP spin hypotheses considered in this work.} 
    \label{tab:hyp}
\end{table}

\begin{figure*}[t]
\begin{center}
\begin{minipage}[t]{0.475\linewidth}
\centering
\includegraphics[width=\textwidth]{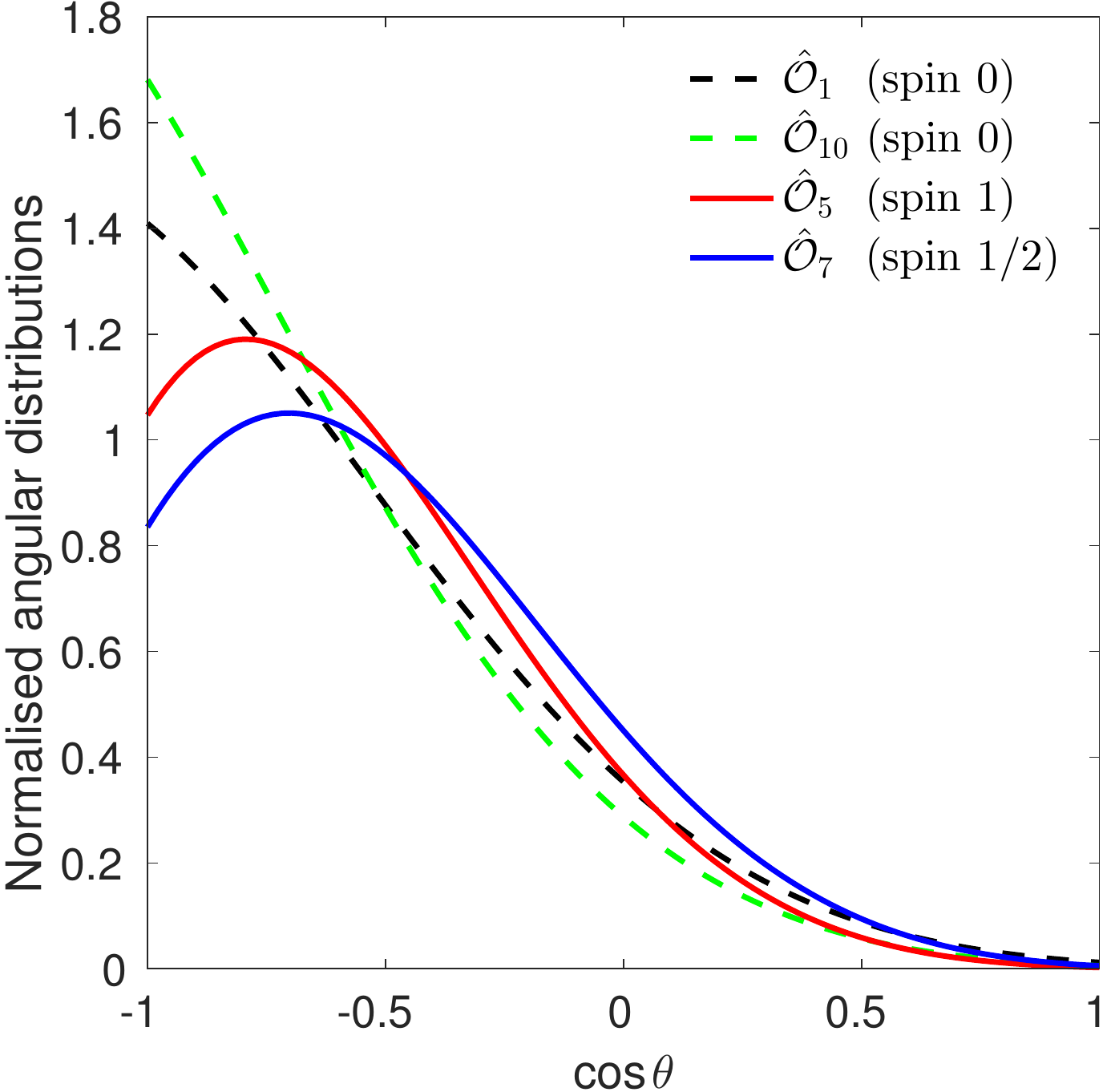}
\end{minipage}
\begin{minipage}[t]{0.491\linewidth}
\centering
\includegraphics[width=\textwidth]{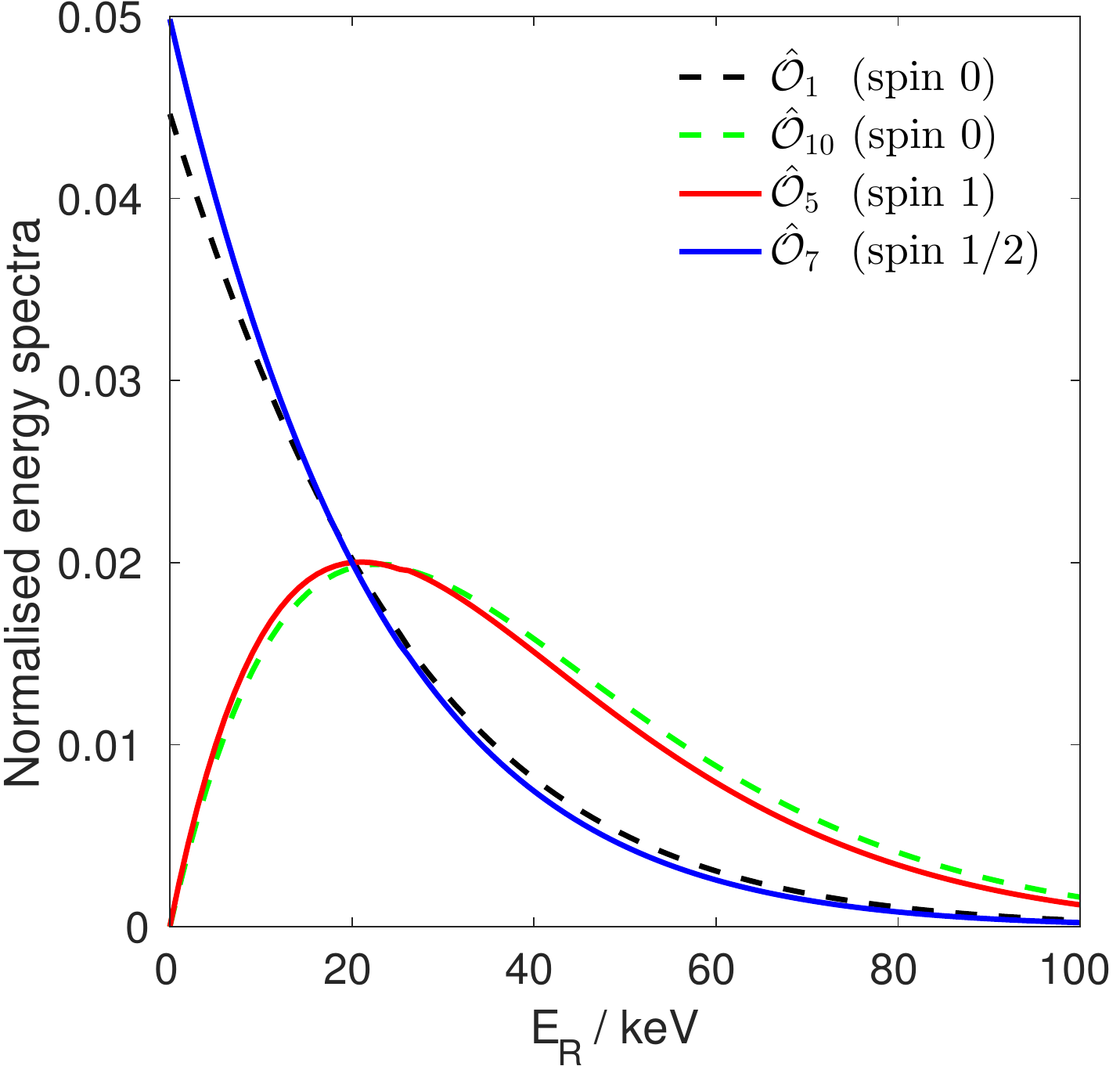}
\end{minipage}
\end{center}
\caption{\textbf{Left panel}.~Normalised angular distribution of signal events as a function of the nuclear recoil direction for selected WIMP-nucleon interactions.~For spin 0 WIMPs, most of the nuclear recoils are expected at $\cos\theta=-1$, i.e., in a direction opposite to the Earth's motion in the galactic rest frame.~For WIMP-nucleon interactions specific to spin 1 or spin 1/2 dark matter, most of the nuclear recoil events are expected in rings around $\cos\theta=-1$.~\textbf{Right panel}.~Normalised energy spectrum of signal events as a function of the nuclear recoil energy for selected WIMP-nucleon interactions.}
\label{fig:met}
\end{figure*}

\subsection{Directional detection}
In this section we briefly review the basic concepts of directional dark matter detection, linking the differential cross-section for WIMP-nucleus scattering to the observable rate of nuclear recoil events.

The angular distribution of WIMP-induced nuclear recoils in low-background detectors is expected to be anisotropic.~Anisotropies are expected since WIMPs preferentially reach the Earth from a direction opposite to Earth's motion in the galactic rest frame~\cite{Spergel:1987kx}.~Dark matter directional detectors have been designed to search for anisotropy patterns in the sphere of nuclear recoil directions.~The double differential rate of nuclear recoil events per unit detector mass is given by
\begin{eqnarray}
\label{eq:rate}
\frac{{\rm d}^2\mathcal{R}}{{\rm d}E_R\,{\rm d}\Omega} =  \kappa_\chi 
 \int {\rm d}^3{\bf v} \, \delta({\bf v}\cdot {\bf w}-w_{q})\, f({\bf v} + {\bf
 v_\oplus}(t)) v^2 \frac{{\rm d}\sigma}{{\rm d}E_R} \nonumber \\
\end{eqnarray}
for $(w_q+|{\bf v_\oplus}|\cos\theta)<v_{\rm esc}$ and zero otherwise.~In Eq.~(\ref{eq:rate}),  $\kappa_\chi=\rho_\chi/(2\pi\,m_\chi m_{A})$, $\rho_\chi\simeq0.4$~GeV~cm$^{-3}$ is the local dark matter density~\cite{Catena:2009mf}, ${\bf w}$ is a unit vector pointing toward the direction of nuclear recoil, $w_{q}=q/(2\mu_{\chi A})$ is the minimum velocity $|\mathbf{v}|=v$  accessible in the scattering, $m_A$ is the target nucleus mass, and ${\bf v_\oplus}(t)$ is the time-dependent Earth's velocity in the galactic rest frame.~For the differential cross-section ${\rm d}\sigma/{\rm d}E_R$, we use Eqs.~(\ref{eq:sigma1}), (\ref{eq:sigma1/2}) or (\ref{eq:sigma0}), depending on the WIMP spin.~Assuming azimuthal symmetry around the direction of ${\bf v_\oplus}(t)$, ${\rm d}\Omega=2\pi {\rm d}\hspace{-0.5mm}\cos\theta$.~The angle $\theta$ is measured with respect to ${\bf v_\oplus}(t)$.~For the velocity distribution of WIMPs in the halo, $f({\bf v} + {\bf v_\oplus}(t))$, we assume a Gaussian function truncated at the escape velocity $v_{\rm esc}=533$~km~s$^{-1}$.~We set the local standard of rest velocity to 220~km~s$^{-1}$~\cite{Catena:2011kv,Bozorgnia:2013pua}.~In the simulations of Sec.~\ref{sec:met}, we focus on hypothetical detectors made of CF$_4$, and use nuclear response functions computed in~\cite{Catena:2015uha}.~Integrating Eq.~(\ref{eq:rate}) over all recoil directions, we find the normalised nuclear recoil energy spectrum 
\begin{align}
\mathcal{P}(E_R) = 2\pi N \int_{|\cos\theta|<1} {\rm d}\hspace{-0.05 cm}\cos\theta\, \left( \frac{{\rm d}^2\mathcal{R}}{{\rm d}E_R\,{\rm d}\Omega}  \right) \,.
\label{eq:ensp}
\end{align}
The constant $N$ is defined by $\int_{E_{\rm th}}^{E_{\rm max}} {\rm d}E_R \,\mathcal{P}(E_R)=1$, where $E_{\rm max}=50$~keV and $E_{\rm th}$ is the energy threshold.~Integrating Eq.~(\ref{eq:rate}) between $E_{\rm th}$ and $50$~keV, we obtain the angular distribution of nuclear recoils
\begin{align}
\mathcal{Q}(\cos\theta) = 2\pi N \int_{E_{\rm th}}^{E_{\rm max}} {\rm d} E_R \, \left( \frac{{\rm d}^2\mathcal{R}}{{\rm d}E_R\,{\rm d}\Omega}  \right) \,.
\label{eq:ansp}
\end{align}
For the threshold $E_{\rm th}$ we consider the benchmark values 5 keV and 20 keV.

Eq.~(\ref{eq:ansp}) peaks at $\cos\theta=-1$ or at $\cos\theta>-1$, depending on whether ${\rm d}\sigma/{\rm d}E_R \propto 1/v^2$ or it scales with $v$ differently~\cite{Catena:2015vpa,Kavanagh:2015jma}.~In the latter case, ring-like features in the sphere of nuclear recoil directions are expected.~If dark matter has spin 0, ${\rm d}\sigma^{(0)}/{\rm d}E_R \propto 1/v^2$, and no rings are expected.~This information can be used to reject the spin 0 WIMP hypothesis in favour of alternative WIMP spin values.~It is important to stress that in this work we focus on ring-like features in $\mathcal{Q}(\cos\theta)$, and not in the double differential rate ${\rm d}^2\mathcal{R}/{\rm d}E_R\,{\rm d}\Omega$.~Ring-like features in ${\rm d}^2\mathcal{R}/{\rm d}E_R\,{\rm d}\Omega$ have been identified previously in an analysis of the $\hat{\mathcal{O}}_1$ operator~\cite{Bozorgnia:2011vc}  or of inelastic exothermic dark matter~\cite{Bozorgnia:2016qkh}.~However, such rings in ${\rm d}^2\mathcal{R}/{\rm d}E_R\,{\rm d}\Omega$ cancel out after integrating over a sufficiently large energy bin, as it is shown below in the left panel of Fig.~\ref{fig:met}, and are therefore not relevant in the context of WIMP spin model selection. 

\section{WIMP spin model selection} 
\label{sec:met}
We now introduce our approach to WIMP spin model selection.~Our approach relies on the framework introduced in Sec.~\ref{sec:theory} and on the hypothetical detection of nuclear recoil events at future directional detection experiments.

\subsection{Methodology}
\label{sec:met2}
Our approach to WIMP spin model selection consists of two stages.~In a first stage, we identify the WIMP-nucleon interaction operators that can be generated -- for a given WIMP spin value -- in the non-relativistic limit of renormalisable single-mediator models for dark matter-quark interactions.~In a second stage, we simulate and analyse data of directional detection experiments to assess whether the identified interactions, and therefore the assumed WIMP spin value, can be rejected in favour of an alternative hypothesis.~In so doing, we also quantify the statistical significance of the rejection.

In this work we focus on the prospects for rejecting the spin 0 WIMP hypothesis, leaving the case of higher WIMP spin values for future studies.~As already mentioned in Sec.~\ref{sec:theory}, $\hat{\mathcal{O}}_1$ and $\hat{\mathcal{O}}_{10}$ are the only operators that can arise from the non-relativistic limit of renormalisable theories for spin 0 dark matter.~Therefore, rejecting the spin 0 WIMP hypothesis is equivalent to rejecting the interaction operators $\hat{\mathcal{O}}_1$ and $\hat{\mathcal{O}}_{10}$ in favour of alternative interactions.~Alternative interactions considered in the present work are $\hat{\mathcal{O}}_5$, which is specific to spin 1 dark matter, and $\hat{\mathcal{O}}_7$, which can only arise from renormalisable spin 1/2 WIMP models.~The purpose of this work is therefore to compare nuclear recoil energy spectra and angular distributions generated by $\hat{\mathcal{O}}_1$, $\hat{\mathcal{O}}_{10}$ and $\hat{\mathcal{O}}_{7}$ ($\hat{\mathcal{O}}_{5}$) with data from next generation directional detection experiments, and through this comparison assess under which conditions the spin 0 WIMP hypothesis can be rejected in favour of the alternative spin 1/2 (1) WIMP hypothesis.

The feasibility of our method is illustrated in Fig.~\ref{fig:met}.~The left panel shows the normalised angular distribution of recoil events, $\mathcal{Q}(\cos\theta)$, for selected WIMP-nucleon interactions.~$\mathcal{Q}(\cos\theta)$ significantly depends on the dark matter particle spin.~If WIMPs have spin 0, most of the nuclear recoils are expected at $\cos\theta=-1$, i.e., in a direction opposite to the Earth's motion in the galactic rest frame.~In contrast, for WIMP-nucleon interactions specific to spin 1 or spin 1/2 dark matter, such as for example the interactions $\hat{\mathcal{O}}_5$ and $\hat{\mathcal{O}}_7$, most of the nuclear recoil events are expected in rings around $\cos\theta=-1$.~Therefore, the search for ring-like features in the sphere of nuclear recoil directions can be used as a tool to reject the hypothesis of spin 0 dark matter.~The right panel in Fig.~\ref{fig:met} shows the normalised energy spectrum $\mathcal{P}(E_R)$ of selected WIMP-nucleon interactions.~The energy spectra of the operators $\hat{\mathcal{O}}_1$ and $\hat{\mathcal{O}}_7$ have similar shapes.~The same is true for the energy spectra of the operators $\hat{\mathcal{O}}_5$ and $\hat{\mathcal{O}}_{10}$.

\begin{figure*}[t]
\begin{center}
\begin{minipage}[t]{0.49\linewidth}
\centering
\includegraphics[width=\textwidth]{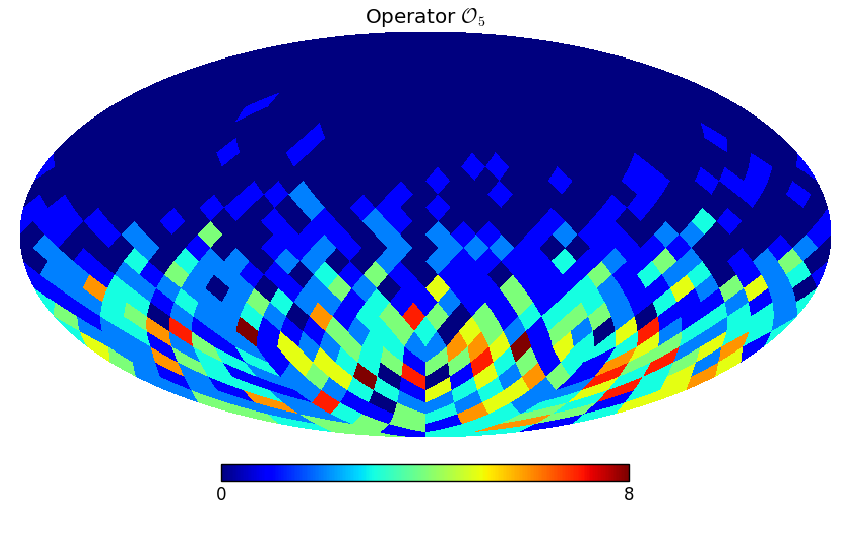}
\end{minipage}
\begin{minipage}[t]{0.49\linewidth}
\centering
\includegraphics[width=\textwidth]{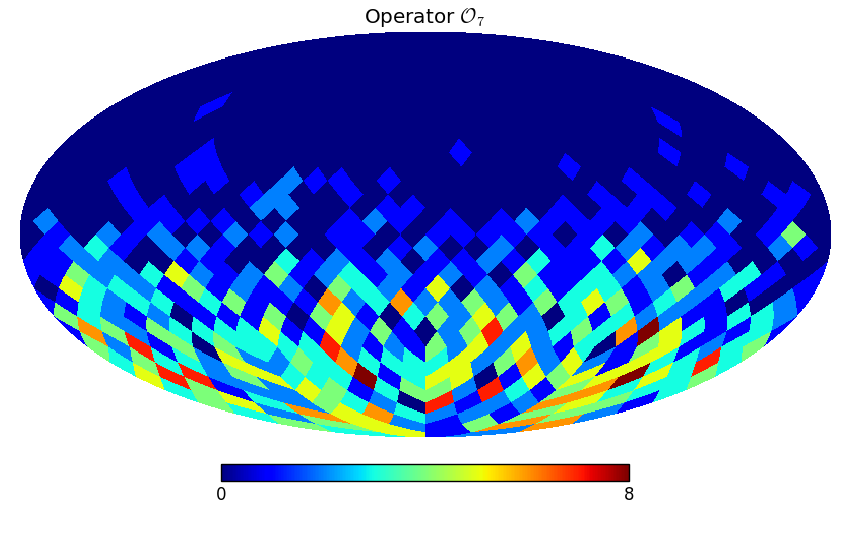}
\end{minipage}
\end{center}
\caption{Mollweide projection of 1000 nuclear recoil events simulated under the $\mathscr{H}_5^{s=1}$ (left panel) and $\mathscr{H}_7^{s=1/2}$ (right panel) hypotheses.~The sphere of nuclear recoil direction has been discretised according to HEALPix's pixelization scheme.~In the simulation we have assumed $N_{\rm pix}=768$ pixels.~The color code follows the number of recoil events per pixel.}  
\label{fig:mp}
\end{figure*}

\subsection{Simulations}
\label{sec:sim}
For simplicity, from now onwards we adopt the notation introduced in Sec.~\ref{sec:theory}, Tab.~\ref{tab:hyp}, when referring to model hypotheses.

In order to validate the method for WIMP spin model selection that we have proposed in Sec.~\ref{sec:met2}, and assess the prospects for rejecting the spin 0 WIMP hypothesis with directional detectors, we use simulated data.~Data are separately simulated under the hypotheses that we would like to reject, $\mathscr{H}_1^{s=0}$ and $\mathscr{H}_{10}^{s=0}$, and under the alternative hypotheses $\mathscr{H}_5^{s=1}$ and $\mathscr{H}_7^{s=1/2}$.~In each simulation, the WIMP mass is set to $m_\chi=100$~GeV, and the coupling constants $c_1$, $c_{10}$, $c_5$ or $c_7$ to the constant $c$.~The constant $c$ is fixed by the requirement $N_{\rm th}(m_\chi,c)=N_{\rm S}$, where $N_{\rm S}$ is the desired number of signal events, and $N_{\rm th}$ the corresponding theoretical expectation (calculable from $\mathcal{P}(E_R)$ in Eq.~(\ref{eq:ensp})).~The total number of expected nuclear recoil events, $N_{\rm T}$, is given by $N_{\rm T}=N_{\rm S}+N_{\rm B}$, where $N_{\rm B}$ is the expected number of background events.~In the numerical calculations, we assume $N_{\rm B}=5$ background events uniformly distributed in energy and nuclear recoil directions~\cite{Billard:2009mf}.~Finally, the dark matter particle spin is set to 0, 1/2 or 1, depending on whether data are simulated under the hypothesis $\mathscr{H}_1^{s=0}$ and $\mathscr{H}_{10}^{s=0}$, $\mathscr{H}_7^{s=1/2}$ or $\mathscr{H}_5^{s=1}$, respectively.~Having fixed the model parameters as explained above, nuclear recoil energies and directions are sampled as explained below.~The nuclear recoil energy interval $(E_{\rm th}$ -- 50) keV is divided in $N_{\rm bins}=3$ bins of equal size.~The observed number of events in the $j$-th energy bin, $m_{j}$, is sampled from a Poisson distribution of mean equal to the number of expected events in that bin, $M_{{\rm T},j}$.~Signal and background events contribute to $M_{{\rm T},j}$.~For simplicity, here we assume perfect energy resolution.~The sphere of nuclear recoil directions is discretised according to HEALPix's pixelization scheme~\cite{Gorski:2004by}.~We assume $N_{\rm pix}=768$ pixels, corresponding to an angular resolution of about 15$^\circ$ (FWHM)~\cite{Billard:2009mf}.~The observed number of nuclear recoils in the $i$-th pixel, $n_{i}$, is sampled from a Poisson distribution of mean equal to the number of expected nuclear recoils in that pixel, $N_{{\rm T},i}$.~Signal and background events contribute to $N_{{\rm T},i}$.~As an illustrative example of our simulations, the Mollweide projection of 1000 nuclear recoil events simulated under the $\mathscr{H}_5^{s=1}$ ($\mathscr{H}_7^{s=1/2}$) hypothesis is reported in left (right) panel of Fig~\ref{fig:mp}.
\begin{figure*}[t]
\begin{center}
\begin{minipage}[t]{0.49\linewidth}
\centering
\includegraphics[width=\textwidth]{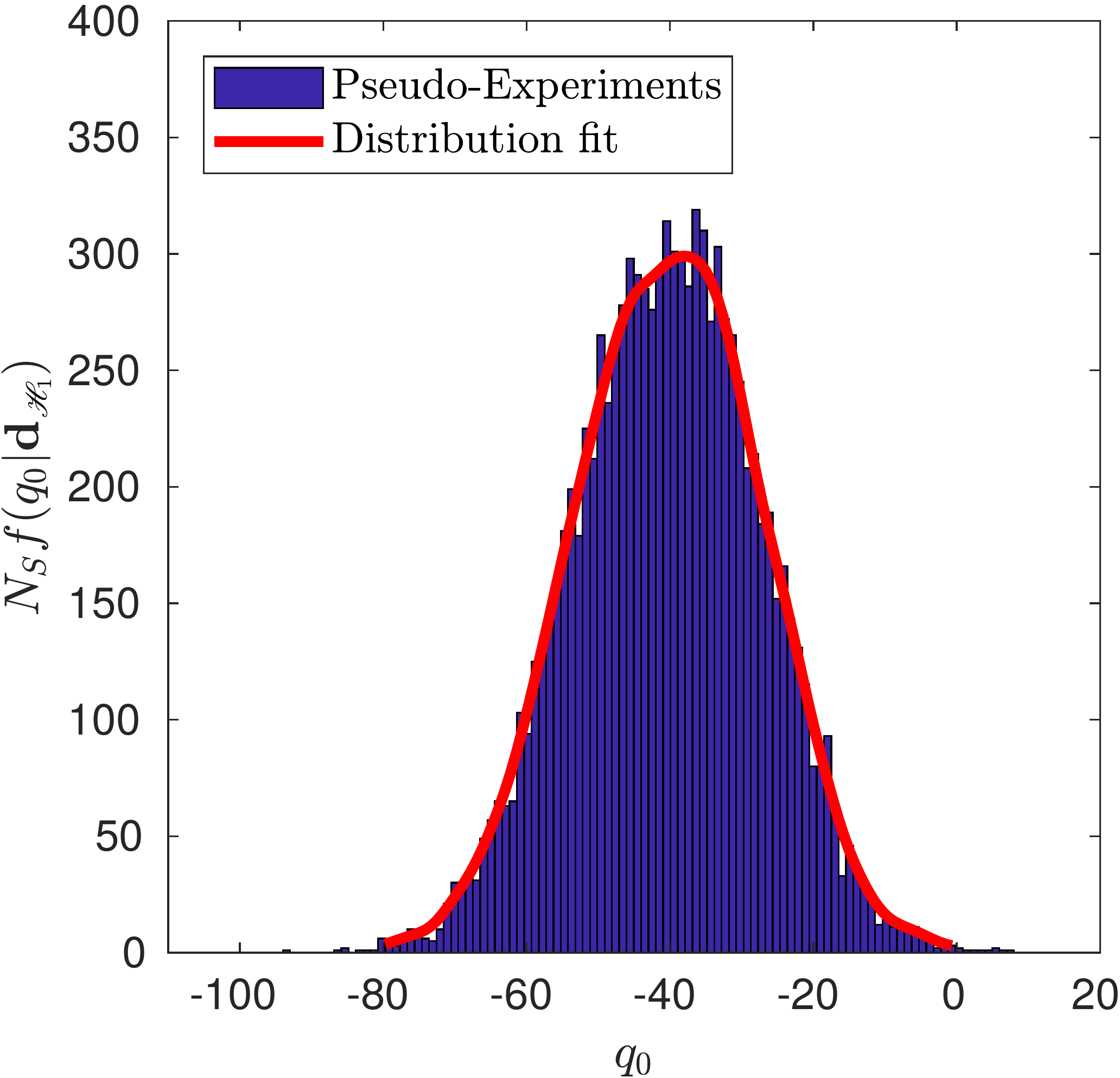}
\end{minipage}
\begin{minipage}[t]{0.493\linewidth}
\centering
\includegraphics[width=\textwidth]{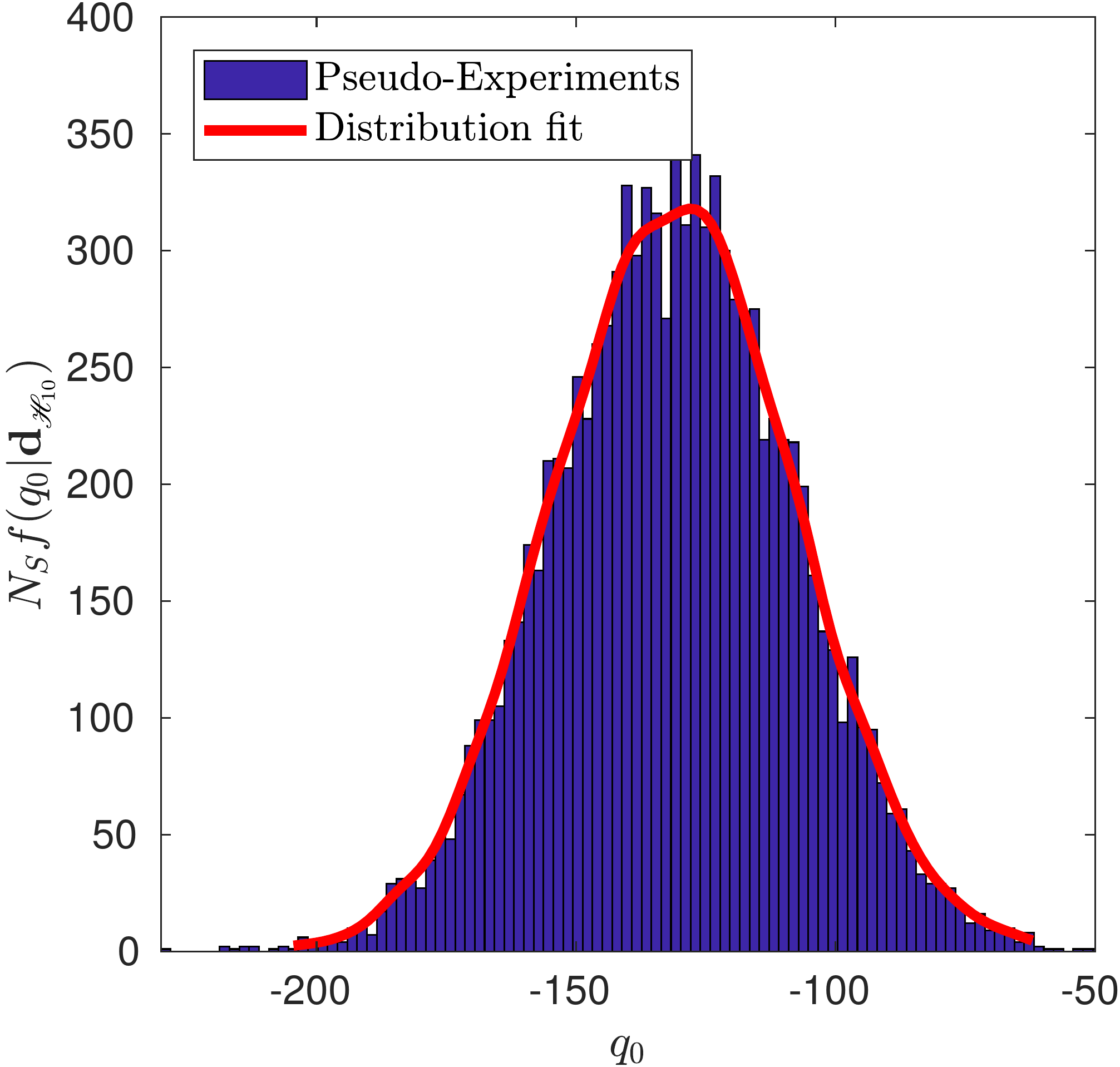}
\end{minipage}
\end{center}
\caption{Probability density functions times $N_S$: $N_S f(q_0|\boldsymbol{d}_{\mathscr{H}_1^{s=0}})$ in left panel and $N_S f(q_0|\boldsymbol{d}_{\mathscr{H}_{10}^{s=0}})$ in the right panel.~Histograms have been obtained from 10000 pseudo-experiments characterised by $N_S=1000$ and $\mathscr{H}_A=\mathscr{H}_7^{s=1/2}$.~Nuclear recoil energies and directions have been simulated as explained in Sec.~\ref{sec:met}.~For illustrative purposes, a fitting distribution has been superimposed to the histograms.}  
\label{fig:h}
\end{figure*}

\subsection{Analysis}
\label{sec:an}
Our goal is to calculate the statistical significance with which the hypotheses $\mathscr{H}_1^{s=0}$ and $\mathscr{H}_{10}^{s=0}$ can be rejected when nuclear recoil energies and directions are generated under $\mathscr{H}_5^{s=1}$ or $\mathscr{H}_7^{s=1/2}$.~We express the statistical significance of the rejection in terms of $p$-values, and present results as a function of the number of signal events $N_S$.~When $\mathscr{H}_1^{s=0}$ and $\mathscr{H}_{10}^{s=0}$ can be rejected with a given statistical significance, the spin 0 WIMP hypothesis can be rejected with the same statistical significance.~Our calculations are based upon the following test statistics 
\begin{align}
q_0=-2\ln\left[\frac{\mathscr{L}(\boldsymbol{d}\,|\,\widehat{\boldsymbol{\Theta}}_B,\mathscr{H}_B)}{\mathscr{L}(\boldsymbol{d}\,|\,\widehat{\boldsymbol{\Theta}}_A,\mathscr{H}_A)}\right] \,,
\label{eq:q0}
\end{align}
where $\mathscr{L}$ is the Likelihood function of the simulated data $\mathbf{d}$, $\boldsymbol{\Theta}_A=(m_\chi, c_A)$ and $\boldsymbol{\Theta}_B=(m_\chi, c_B)$ are the dark matter particle mass and coupling constants characterising the hypotheses $\mathscr{H}_A$ and $\mathscr{H}_B$, respectively, and $\widehat{\boldsymbol{\Theta}}_A$ ($\widehat{\boldsymbol{\Theta}}_B$) is the value of $\boldsymbol{\Theta}_A$ ($\boldsymbol{\Theta}_B$) that maximises the Likelihood $\mathscr{L}$ when fitting the data $\boldsymbol{d}$ under the hypothesis $\mathscr{H}_A$ ($\mathscr{H}_B$).~Specifically, here we are interested in the four scenarios:
\begin{enumerate}
\item $\mathscr{H}_A=\mathscr{H}_5^{s=1}$, $\mathscr{H}_B=\mathscr{H}_1^{s=0}$; 
\item $\mathscr{H}_A=\mathscr{H}_5^{s=1}$, $\mathscr{H}_B=\mathscr{H}_{10}^{s=0}$; 
\item $\mathscr{H}_A=\mathscr{H}_7^{s=1/2}$, $\mathscr{H}_B=\mathscr{H}_1^{s=0}$; 
\item $\mathscr{H}_A=\mathscr{H}_7^{s=1/2}$, $\mathscr{H}_B=\mathscr{H}_{10}^{s=0}$\,.
\end{enumerate}
Accordingly, the coupling constants $c_A$ and $c_B$ in Eq.~(\ref{eq:q0}) can be $c_1$, $c_{10}$, $c_5$ or $c_7$, depending on the scenario in analysis.~For $\mathscr{L}$, we assume the product of Poisson distributions 
\begin{widetext}
\begin{align}
\mathscr{L}(\boldsymbol{d}\,|\,\boldsymbol{\Theta},\mathscr{H})= \prod_{i=1}^{N_{\rm pix}} \prod_{j=1}^{N_{\rm bins}} \frac{N_{{\rm T},i}(\boldsymbol{\Theta},\mathscr{H})^{n_i(\boldsymbol{d})}}{n_i(\boldsymbol{d})!} \frac{M_{{\rm T},j}(\boldsymbol{\Theta},\mathscr{H})^{m_j(\boldsymbol{d})}}{m_j(\boldsymbol{d})!} \,e^{-\left[N_{{\rm T},i}(\boldsymbol{\Theta},\mathscr{H})+M_{{\rm T},j}(\boldsymbol{\Theta},\mathscr{H})\right]}\,,
\end{align}
\end{widetext}
where the pair ($\boldsymbol{\Theta},\mathscr{H}$) can be one of the four combinations listed above.

Given the test statistics in Eq.~(\ref{eq:q0}), the statistical significance with which $\mathscr{H}_1^{s=0}$ and $\mathscr{H}_{10}^{s=0}$ can be rejected is computed as follows.~For each value of $N_{\rm S}$ that we consider, we simulate 1000 pseudo-experiments under the hypothesis $\mathscr{H}_A$, i.e.~$\mathscr{H}_5^{s=1}$ or $\mathscr{H}_7^{s=1/2}$.~We then construct the probability density function (PDF) of $q_0$ under $\mathscr{H}_A$, $f(q_0|\boldsymbol{d}_{\mathscr{H}_A})$, and calculate the associated median, $q_{\rm med}$.~$q_{\rm med}$ represents the ``typical'' value of $q_0$ when WIMPs interacts according to $\mathscr{H}_A$.~Subsequently, we simulate 10000 pseudo-experiments under the hypothesis $\mathscr{H}_B$, i.e.~$\mathscr{H}_1^{s=0}$ or $\mathscr{H}_{10}^{s=0}$.~From these simulations we obtain the PDF of $q_0$ under $\mathscr{H}_B$, $f(q_0|\boldsymbol{d}_{\mathscr{H}_B})$, and calculate the associated $p$-value:
\begin{align}
p=\int_{q_{\rm med}}^\infty{\rm d}q_0 \,f(q_0|\boldsymbol{d}_{\mathscr{H}_B})\,.
\label{eq:p}
\end{align}
The $p$-value in Eq.~(\ref{eq:p}) is our measure of the statistical significance with which the hypotheses $\mathscr{H}_1^{s=0}$ and $\mathscr{H}_{10}^{s=0}$ can be rejected as a function of $N_S$.~For $N_S=1000$ and $\mathscr{H}_A=\mathscr{H}_7^{s=1/2}$, the PDFs $f(q_0|\boldsymbol{d}_{\mathscr{H}_1^{s=0}})$ and $f(q_0|\boldsymbol{d}_{\mathscr{H}_{10}^{s=0}})$ are reported in the left and right panels of Fig.~\ref{fig:h}, respectively. 

\begin{figure*}[t]
\begin{center}
\begin{minipage}[t]{0.49\linewidth}
\centering
\includegraphics[width=\textwidth]{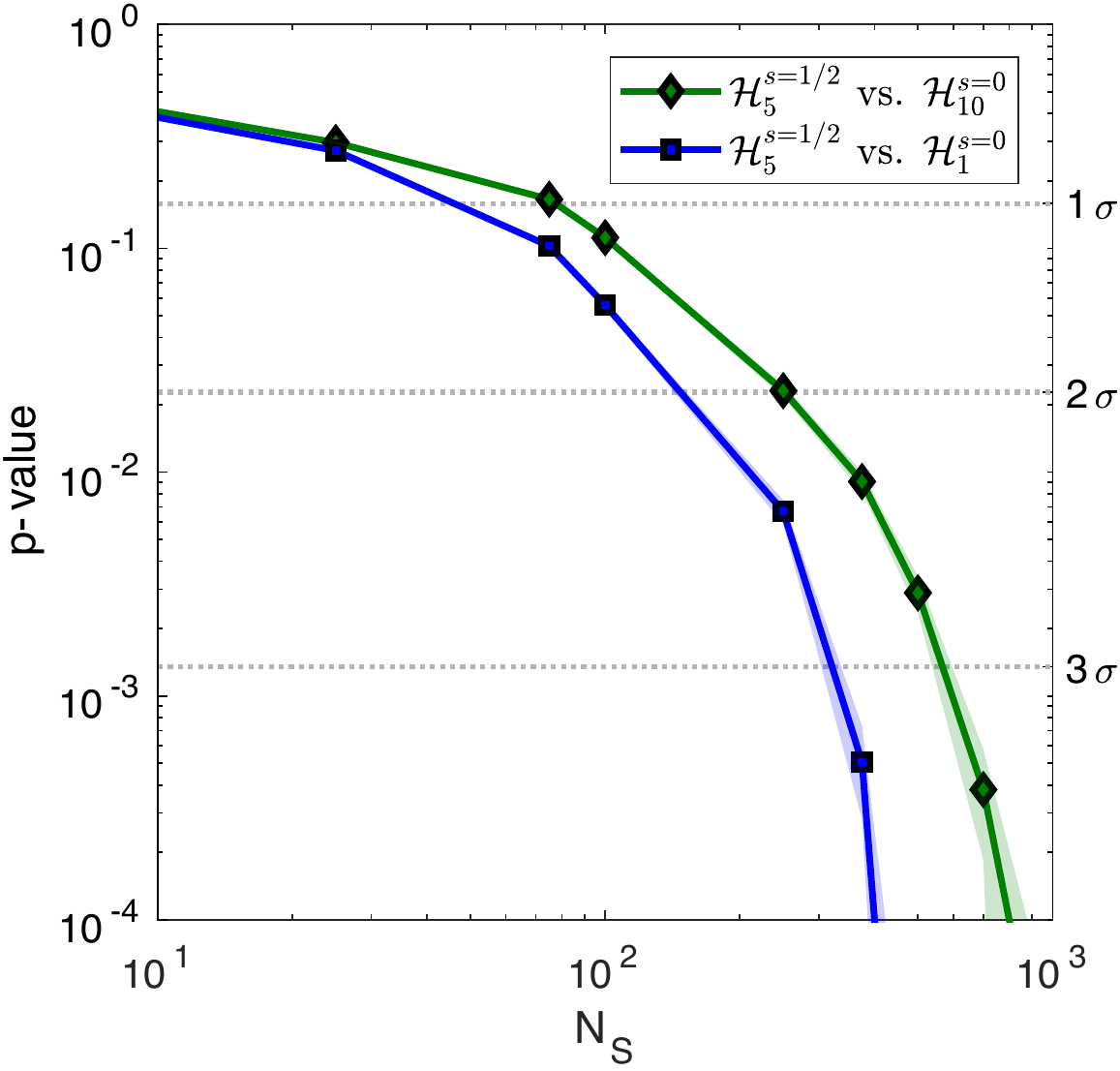}
\end{minipage}
\begin{minipage}[t]{0.49\linewidth}
\centering
\includegraphics[width=\textwidth]{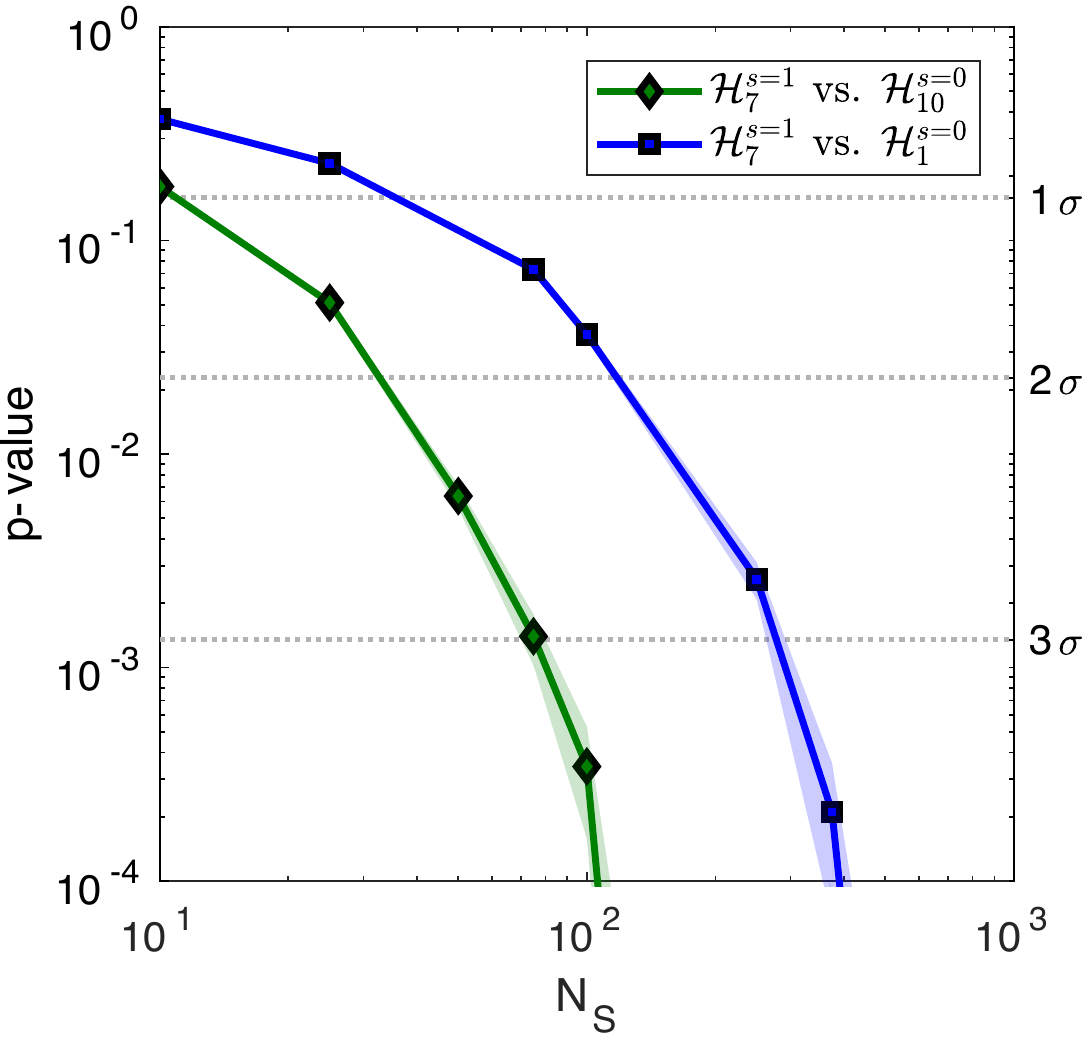}	
\end{minipage}
\end{center}
\caption{Statistical significance with which nuclear recoil energies and directions, $\boldsymbol{d}$, simulated under the hypothesis $\mathscr{H}_5^{s=1}$ (left panel) and $\mathscr{H}_7^{s=1/2}$ (right panel) allow to reject the hypotheses $\mathscr{H}_1^{s=0}$ (blue curve) and $\mathscr{H}_{10}^{s=0}$ (green curve)  as a function of the number of signal events $N_S$.~For values of $N_S$ such that both $\mathscr{H}_1^{s=0}$ and $\mathscr{H}_{10}^{s=0}$ can be rejected, the spin 0 WIMP hypothesis can also be rejected.~To each curve in the figure, we have associated shaded binomial error bands, i.e.~$\Delta p=\sqrt{p(1-p)/N}$, with $N=10000$.}
\label{fig:p}
\end{figure*}

\section{Results}
\label{sec:results}
From Eq.~(\ref{eq:p}), we now calculate as a function of $N_S$ the $p$-values corresponding to the four already mentioned scenarios: 
\begin{enumerate}
\item $\mathscr{H}_A=\mathscr{H}_5^{s=1}$, $\mathscr{H}_B=\mathscr{H}_1^{s=0}$; 
\item $\mathscr{H}_A=\mathscr{H}_5^{s=1}$, $\mathscr{H}_B=\mathscr{H}_{10}^{s=0}$; 
\item $\mathscr{H}_A=\mathscr{H}_7^{s=1/2}$, $\mathscr{H}_B=\mathscr{H}_1^{s=0}$; 
\item $\mathscr{H}_A=\mathscr{H}_7^{s=1/2}$, $\mathscr{H}_B=\mathscr{H}_{10}^{s=0}$\,.
\end{enumerate}
Through this calculation, we quantitatively assess the prospects for rejecting the spin 0 WIMP hypothesis at next generation directional detection experiments.~Our default choices for energy threshold and dark matter particle mass are $E_{\rm th}=20$~keV and $m_{\chi}=100$~GeV, respectively.~A lower energy threshold, i.e.~$E_{\rm th}=5$~keV, is considered in Sec.~\ref{sec:low}.

\subsection{Spin 1 vs. spin 0 hypothesis}
\label{sec:10}
In this first application of our method for WIMP spin model selection, we aim at rejecting the spin 0 WIMP hypothesis in favour of the alternative hypothesis $\mathscr{H}_A=\mathscr{H}_5^{s=1}$, according to which dark matter has spin 1 and interacts with nucleons through the operator $\hat{\mathcal{O}}_5$.~By simulating nuclear recoil energies and directions as explained in Sec.~\ref{sec:sim}, we derive the PDFs of the test statistics $q_0$ in Eq.~(\ref{eq:q0}) under the hypothesis $\mathscr{H}_A=\mathscr{H}_5^{s=1}$, and $\mathscr{H}_B=\mathscr{H}_1^{s=0}$ or $\mathscr{H}_B=\mathscr{H}_{10}^{s=0}$.~From the former PDF we obtain $q_{\rm med}$, from the latter one we calculate the $p$-value of the alternative hypothesis $\mathscr{H}_A=\mathscr{H}_5^{s=1}$ as a function of the number of signal events $N_S$ using Eq.~(\ref{eq:p}).~To each $p$-value that we obtain in this manner, we associate the corresponding statistical significance $Z=\Phi^{-1}(1-p)$ expressed in units of the standard deviation $\sigma$ of a fictitious Gaussian distribution $\Phi$~\cite{Cowan:2010js}.~Results for this calculation are reported in the left panel of Fig.~\ref{fig:p}.~This figure shows the $p$-value and associated statistical significance of the alternative hypothesis $\mathscr{H}_A=\mathscr{H}_5^{s=1}$ as a function of the number of signal events $N_S$.~The blue curve corresponds to the case $\mathscr{H}_{B}=\mathscr{H}_1^{s=0}$, whereas the green curve has been obtained for $\mathscr{H}_{B}=\mathscr{H}_{10}^{s=0}$.~A 2$\sigma$ rejection of the $\mathscr{H}_1^{s=0}$ hypothesis requires about 150 signal events, whereas about 250 signal events are needed in order to reject the hypothesis $\mathscr{H}_{10}^{s=0}$ at the same level of statistical significance.~From the results reported in this section, we conclude that, if $E_{\rm th}=20$~keV, about 250 signal events are needed in order to reject the spin 0 WIMP hypothesis with a statistical significance of 2$\sigma$.

\begin{figure*}[t]
\begin{center}
\begin{minipage}[t]{0.49\linewidth}
\centering
\includegraphics[width=\textwidth]{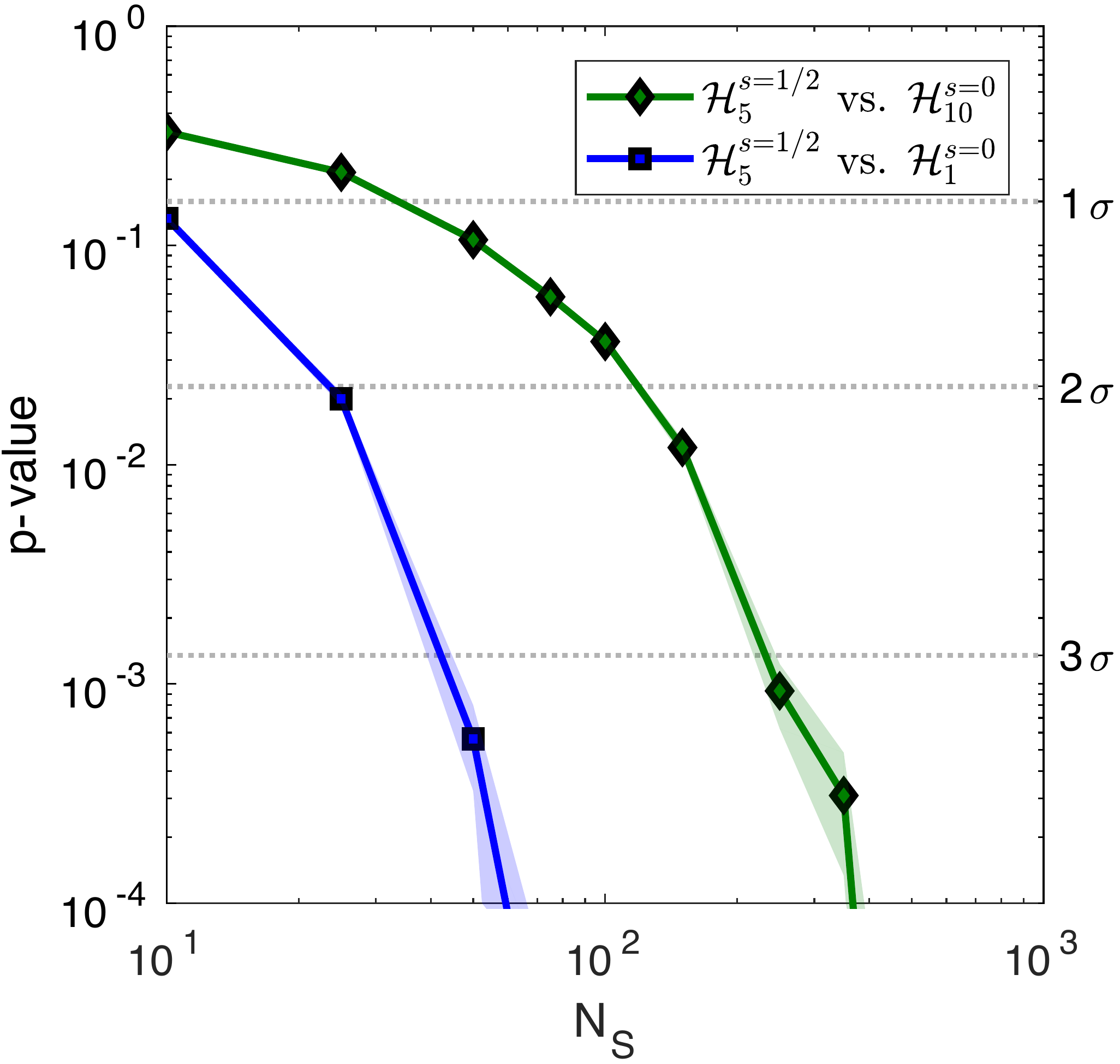}
\end{minipage}
\begin{minipage}[t]{0.49\linewidth}
\centering
\includegraphics[width=\textwidth]{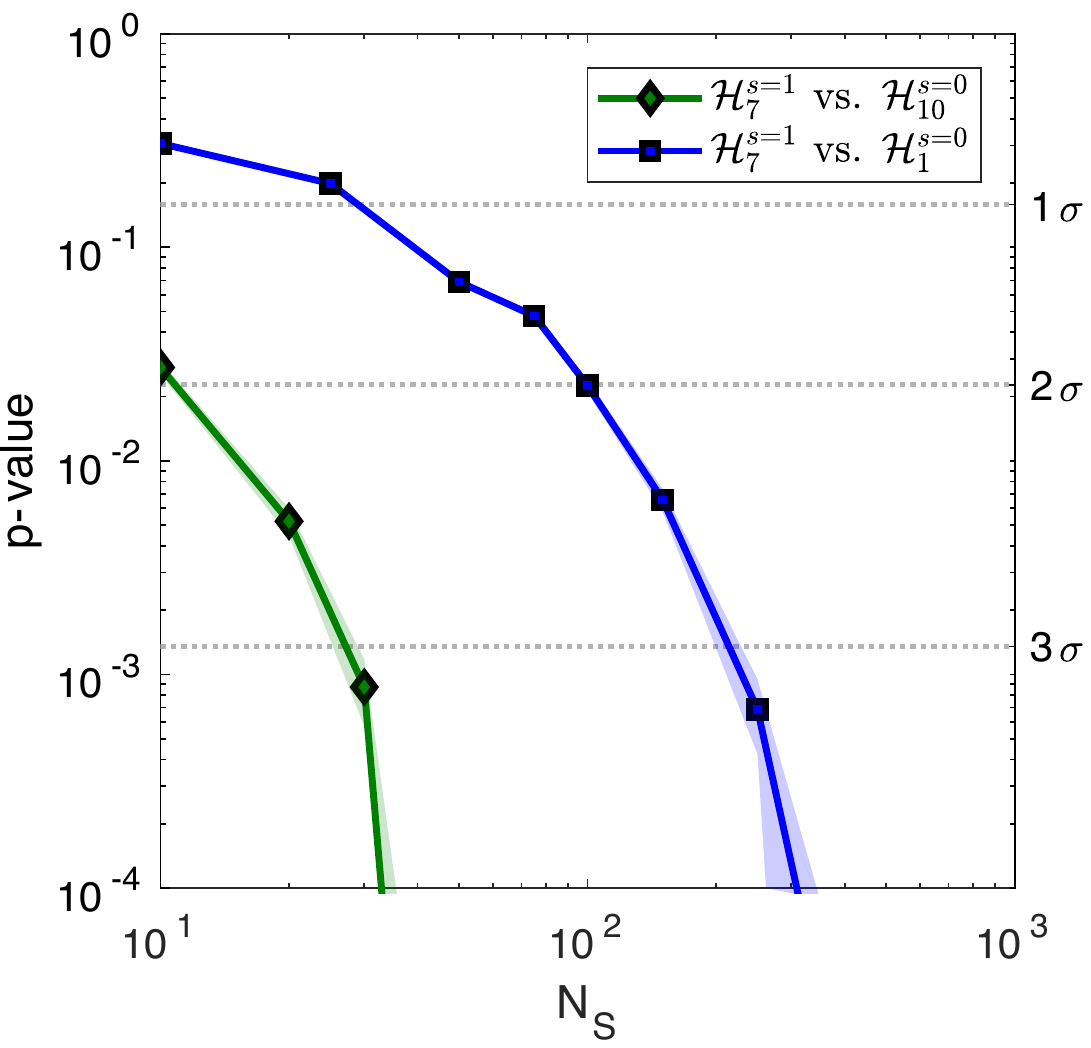}
\end{minipage}
\end{center}
\caption{Same as for Fig.~\ref{fig:p}, but now with $E_{\rm th}=5$~keV instead of $E_{\rm th}=20$~keV.}  
\label{fig:plow}
\end{figure*}

\subsection{Spin 1/2 vs. spin 0 hypothesis}
\label{sec:1/20}
In this subsection, we aim at rejecting the spin 0 WIMP hypothesis in favour of the alternative hypothesis $\mathscr{H}_A=\mathscr{H}_7^{s=1/2}$, according to which dark matter has spin 1/2 and interacts with nucleons through the operator $\hat{\mathcal{O}}_7$.~Following the same procedure already illustrated in Sec.~\ref{sec:an}, we compute the $p$-value and associated statistical significance of the hypothesis $\mathscr{H}_7^{s=1/2}$, separately considering $\mathscr{H}_B=\mathscr{H}_1^{s=0}$ and $\mathscr{H}_B=\mathscr{H}_{10}^{s=0}$.~The right panel in Fig.~\ref{fig:p} shows the $p$-values obtained from this calculation as a function of the number of signal events, $N_S$.~As for the left panel in the same figure, the blue curve corresponds to the case $\mathscr{H}_{B}=\mathscr{H}_1^{s=0}$, whereas the green curve has been obtained for $\mathscr{H}_{B}=\mathscr{H}_{10}^{s=0}$.~Also in this case, the $y$-axis reports the number of standard deviations, $\sigma$, associated with a given $p$-value.~A 2$\sigma$ rejection of the $\mathscr{H}_1^{s=0}$ hypothesis requires about 150 signal events.~At the same time, about 30 signal events are needed in order to reject the hypothesis $\mathscr{H}_{10}^{s=0}$ at the same level of statistical significance.~Compared to the case $\mathscr{H}_A=\mathscr{H}_5^{s=1}$, less signal events are needed in order to reject the spin 0 WIMP hypothesis in favour of the alternative hypothesis $\mathscr{H}_A=\mathscr{H}_7^{s=1/2}$.~This result is expected, and related to the shape of $\mathcal{Q}(\cos\theta)$, which peaks at values of $\cos\theta$ significantly different from -1 for $\mathscr{H}_A=\mathscr{H}_7^{s=1/2}$.

\subsection{Lowering the energy threshold}
\label{sec:low}
Results presented in the previous subsections assume $E_{\rm th}=20$~keV.~Because of this relatively large energy threshold, energy information played a secondary role in the previous analyses, as one can also see from the right panel in Fig.~\ref{fig:met}.~In this subsection, we recalculate the $p$-values reported in Fig.~\ref{fig:p}, now with the lower energy threshold, $E_{\rm th}=5$~keV.~Performing WIMP spin model selection, a lower threshold energy allows to exploit differences in the recoil energy spectra that are highlighted in the right panel of Fig.~\ref{fig:met}.~Results of our low-threshold analysis are reported in Fig.~\ref{fig:plow}.~By lowering the energy threshold and accounting for energy information, we find a significant decrease in the $p$-value, which implies a more effective rejection of the spin 0 WIMP hypothesis.~For instance, when data are generated under the $\mathscr{H}_7^{s=1/2}$ hypothesis, a 2$\sigma$ rejection of the $\mathscr{H}_1^{s=0}$ hypothesis requires $N_S\simeq$100, and $N_S\simeq$10 signal events are needed in order to reject the hypothesis $\mathscr{H}_{10}^{s=0}$ with a statistical significance at the 2$\sigma$ level.~At the same time, when data are generated under the $\mathscr{H}_5^{s=1}$ hypothesis, a 2$\sigma$ rejection of the $\mathscr{H}_1^{s=0}$ ($\mathscr{H}_{10}^{s=0}$) hypothesis requires $N_S\simeq$25 ($N_S\simeq$150).

\section{Conclusion}
\label{sec:conclusion}
We have computed the number of signal events needed to reject the hypothesis of spin 0 WIMP at next generation directional detection experiments exploiting CF$_4$ as a target material.~Assuming an energy threshold of 5 keV, we have found that about 100 nuclear recoils will be enough to enable a $2\sigma$ rejection of the spin 0 dark matter hypothesis in favour of alternative hypotheses where dark matter has spin 1 or 1/2.~For comparison, about 100 signal events are expected in a CF$_4$ detector operating at a pressure of 30 torr with an exposure of approximately 26,000 cubic-meter-detector days for WIMPs of 100 GeV mass and a WIMP-Fluorine scattering cross-section of 0.25~pb.~Interestingly, the DMTPC collaboration has shown that comparable exposures are within reach of an array of cubic meter TPC detectors.~\cite{Deaconu:2017vam}.~Our results are based upon the following considerations.~If the dark matter particle has spin 0, only the operators $\hat{\mathcal{O}}_1$ and $\hat{\mathcal{O}}_{10}$ can arise from the non-relativistic reduction of renormalisable single-mediator models for dark matter-quark interactions.~The operators $\hat{\mathcal{O}}_1$ and $\hat{\mathcal{O}}_{10}$ generate angular distributions of nuclear recoil events at directional detection experiments that have a maximum in the direction antiparallel to the Earth's motion in the galactic rest frame.~In contrast, interaction operators specific to spin 1 and spin 1/2 dark matter can generate angular distributions of nuclear recoil events that peak in rings centred around the direction of the Earth's motion.~Following~\cite{Fitzpatrick:2012ib}, we have denoted these interaction operators by $\hat{\mathcal{O}}_5$ and $\hat{\mathcal{O}}_7$, respectively.~Here we have shown that ring-like features in the sphere of nuclear recoil directions are potentially observable at next generation directional detection experiments, and can therefore be used to reject the spin 0 WIMP hypothesis.

For completeness, we also provide examples of scenarios in which our method for WIMP spin model selection cannot be applied.~In this context, we would like to stress that the operators $\hat{\mathcal{O}}_1$ and $\hat{\mathcal{O}}_{10}$ can also be generated if dark matter has spin 1/2 or 1~\cite{Dent:2015zpa}.~Therefore, if dark matter has spin different from zero and interacts with nucleons through the operators $\hat{\mathcal{O}}_1$ and $\hat{\mathcal{O}}_{10}$, rejecting the spin 0 WIMP hypothesis on the basis of directional detection experiments alone is not possible.~Furthermore, if dark matter has spin different from zero and interacts with nucleons through the operators $\hat{\mathcal{O}}_4$, $\hat{\mathcal{O}}_6$, $\hat{\mathcal{O}}_9$, $\hat{\mathcal{O}}_{11}$ no ring-like features are expected in the sphere of nuclear recoils~\cite{Catena:2015vpa,Kavanagh:2015jma}.~Therefore, rejecting the interactions $\hat{\mathcal{O}}_1$ and $\hat{\mathcal{O}}_{10}$, i.e.~the spin 0 WIMP hypothesis, would require comparing different spin hypotheses based on energy information only, which might in turn require very large exposures.~Finally, operator evolution might quantitatively affect our conclusions.~For example, it could generate the operator $\hat{\mathcal{O}}_1$ as the leading operator in models where only $\hat{\mathcal{O}}_{10}$ is predicted at tree-level~\cite{DEramo:2014nmf,D'Eramo:2016atc}.~However, since our method for WIMP spin model selection is based upon simultaneously rejecting $\hat{\mathcal{O}}_1$ and $\hat{\mathcal{O}}_{10}$, in these cases operator evolution would not qualitatively change our conclusions.

Let us also comment on the generality of the proposed method for WIMP spin model selection.~Here we have simulated data on nuclear recoil events assuming $\hat{\mathcal{O}}_5$ and $\hat{\mathcal{O}}_7$ as underlying interactions.~Since the operators $\hat{\mathcal{O}}_{8}$, $\hat{\mathcal{O}}_{13}$ and $\hat{\mathcal{O}}_{14}$ give rise to similar ring-like features in the sphere of nuclear recoil directions~\cite{Catena:2015vpa,Kavanagh:2015jma}, the results found in this work also qualitatively apply to the case of spin 1/2 or 1 WIMPs interacting with nucleons through one of the three interaction operators mentioned above.~Finally, we briefly comment on the applicability of the proposed method to the rejection of other WIMP spin values.~For spin 1/2 or 1 WIMPs, a variety of WIMP-nucleon interaction operators can arise from the non-relativistic reduction of renormalisable single-mediator models for dark matter-quark interactions~\cite{Dent:2015zpa}.~Therefore, rejecting the spin 1/2 or 1 WIMP hypothesis appears to be a more difficult task, although in principle possible to address within the systematic approach presented in this work.

\acknowledgments This work has been supported by the Knut and Alice Wallenberg Foundation (PI: Jan Conrad) and is performed in the context of the Swedish Consortium for Direct Detection of Dark Matter (SweDCube).~This research has also been supported by the Munich Institute for Astro- and Particle Physics (MIAPP) within the Deutsche Forschungsgemeinschaft (DFG) cluster of excellence ``Origin and Structure of the Universe''.~Finally, we thank the participants of the MIAPP programme ``Astro-, Particle and Nuclear Physics of Dark Matter Direct Detection'' for many valuable discussions.


%

\end{document}